\documentclass{article}

\usepackage{arxiv}
\usepackage{amsmath}

\usepackage[utf8]{inputenc} % allow utf-8 input
\usepackage[T1]{fontenc}    % use 8-bit T1 fonts
\usepackage{hyperref}       % hyperlinks
\usepackage{url}            % simple URL typesetting
\usepackage{booktabs}       % professional-quality tables
\usepackage{amsfonts}       % blackboard math symbols
\usepackage{nicefrac}       % compact symbols for 1/2, etc.
\usepackage{microtype}      % microtypography
\usepackage{lipsum}		% Can be removed after putting your text content
\usepackage{graphicx}
\usepackage{biblatex}
\usepackage{doi}

\newcommand{\held}{\mathit{held}}

\title{Digital Circuits as Moore Machines}

\date{January 2026} 					% Or removing it

\author{ \href{https://orcid.org/0000-0001-5085-9794} {\includegraphics[scale=0.06]{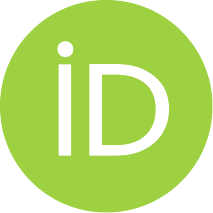}\hspace{1mm}Victor Yodaiken}\thanks{
		Independent Researcher.} \\
	Austin Texas\\
	\texttt{victor.yodaiken@gmail.com} \\
}

\renewcommand{\shorttitle}{\textit{arXiv} PR Sequential}

\hypersetup{
pdftitle={Digital Circuits as Moore Machines}
pdfauthor={Victor Yodaiken},
pdfkeywords={Moore machines, automata, primitive recursive},
}

\begin{document}
\maketitle

\begin{abstract}
Here it is
\end{abstract}

\newcommand{\ess}{\epsilon}
\newcommand{\Set}{\mathit{Set}}
\newcommand{\RReset}{\mathit{Reset}}
% keywords can be removed

\section{Introduction}

The real-time and compositional behavior of digital circuits can be 
represented by state machines using a method for working with large numbers
of states and for interconnecting state machines described in a 
previous paper \cite{yodaikenlarge}. The method treats state machines
as maps from finite sequences of events to the output reached by
following that sequence from the initial state. 
In this case, events represent real-time discrete samples of signals asserted
on the input pins of circuits, sampled at some fixed
frequency. The sampling rate can be left unspecified
or made concrete as needed. 

This note will start with a couple
of simple combinatorial circuits: a bus and nand-gate,
then specify a set-reset latch,
and finally show how connecting two nand-gates can implement the latch.
A remaining section sketches possible alternative approaches using
the same underlying method.

None of this is earthshaking news -  these circuits have 
been widely used and modeled for a long time now and are staples of
introductory digital circuit courses. However, defining the real-time
behavior of these circuits
without the use of a programming language
like framework such as in VHDL, simplifies semantics.
The reader will have to decide for
themselves if there is anything illuminating in representing circuits
as functions.

\section{Circuits}

Let \(\ess\) be the empty sequence and \(w.a\) be the sequence obtained by appending sample \(a\) to \(w\) on the right.
A sample is just a
tuples of binary values:
\[ a = (x_1,\dots x_n)\text{ where } x_i\in \{0,1\}\text{ and }  (a)_i = x_i.\]
The length of an input sequence \(w\) is the amount of time that has
passed since the start state -- in whatever units of time each sample
represents\footnote{All event sequences here are finite as we are
not interested in infinite time periods and infinitisimal 
sample times can be left to standard analytic methods.}.
The state of a circuit state machine
 \(M\) is determined by the sequence  of events
 that has driven the machine from its initial state.
\[ \text{Input Sequence } \Rightarrow \fbox{M} \Rightarrow Outputs\]
The state machines are deterministic but often only partially specified.

For circuit \(C\) and event 
sequence \(w\), \(C(w,r)\) is the output on pin \(r\) in the state reached by following \(w\) from the initial state.
If \(C\) has only one output pin, we can
just write \(C(w)\).

To see how long a signal \(b\in \{0,1\}\) has been applied to 
a particular input pin define:
\[\held(\ess,i,b)= 0\text{ and } \held(w.a,i,b) = \begin{cases}
	(1+\held(w,i,b))&\text{if }(a)_i = b\\
0&\text{otherwise}. \end{cases}\]

Equivalently: 
\[\held(\ess,i,b)= 0\text{ and } \held(w.a,i,b) = ((a)_i=b)(1+\held(w,i,b))\]
where \(((a)_i = b)\) above  has value \(1 \) if true and \(0\) otherwise. 

Then for a bus \(B\) with \(k\) wires, with propagation delay of \(n\) samples: 
\begin{equation} \text{If }\held(w,1,b)\geq n\text{ then }B(w,i)=b\end{equation}

	\newcommand{\G}{\mathtt{G}}
Say \(\G\) is a 2 input \emph{nand-gate} with propagation delay \(n\) if and only if
\begin{equation}
\begin{array}{l} 
\G(w)\in \{0,1\}\\
\G(w)=1 \text{ if }\held(w,1,0) \geq n\text{ or  }\held(w,2,0)\geq n\\
\G(w)=0\text{ if }\held(w,1,1)\geq n\text{ and }\held(w,2,1)\geq n
\end{array}
\end{equation}
\begin{center}  \includegraphics[width=0.3\textwidth]{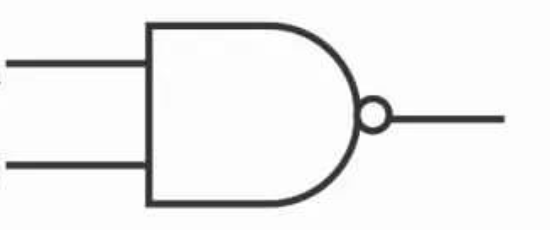} \end{center}
That is, if either input pin is held  low (\(0\) for at least \(n\) samples
the output is \(1\) and if both are held high (\(1\)) for at least \(n\) 
samples the output must be \(0\). If neither condition is met, then all we
know is that the output is either \(0\) or \(1\) (see section 
\ref{sec:alternatives} for other possibilities).

\newcommand{\SR}{\mathtt{SR}}

Now we are going to define a set/reset (SR) latch in two steps. The
remarkable thing about this latch is that when it gets to a stable (latched)
state (set or reset), the inputs  \((1,1)\) keep
it in that state indefinitely.
The bus and nand-gates have a property that there is some \(n\) so that looking back \(n\) samples is sufficient to test e.g. if a signal has been asserted
long enough to determine the output. The latch does not have that property,
an arbitrary sequence of \((1,1)\) won't change the latched value, if
some value has been latched.

\begin{equation}
\begin{array}{l}
\Set(\ess,n) = 0\\
\Set(w.a,n) = \begin{cases}
   1&\text{if }\held(w.a,1,0)\geq n\text{ and }\held(w.a,2,1)\geq n\\
   &\text{or if }\Set(w,n)=1\text{ and }(a)_1= (a)_2 = 1\\
0&\text{otherwise}
\end{cases}\\
\RReset(\ess,n) = 0\\
\RReset(w.a,n) = \begin{cases}
   1&\text{if }\held(w.a,1,1)\geq n\text{ and }\held(w.a,2,0)\geq n\\
   &\text{or if }\RReset(w,n)=1\text{ and }(a)_1= (a)_2 = 1\\
0&\text{otherwise}
\end{cases}
\end{array}
\end{equation}

As usual, input \((0,0)\) is not used. 

\(L\) is a SR latch with latch time \(n\) if and only if
\begin{equation}
\begin{array}{l}
L(w,1) \in \{0,1\}, L(w,2)\in \{0,1\}.\\
\text{If }\RReset(w,n)=1 \text{ then } L(w,1)=1 \text { and } L(w,2)=0\\
\text{If }\Set(w,n)=1 \text{ then } L(w,1)=0 \text { and } L(w,2)=1\\
\end{array}
\end{equation}

Build a latch by cross connecting two 
nand-gates.
\begin{center}  \includegraphics[width=0.3\textwidth]{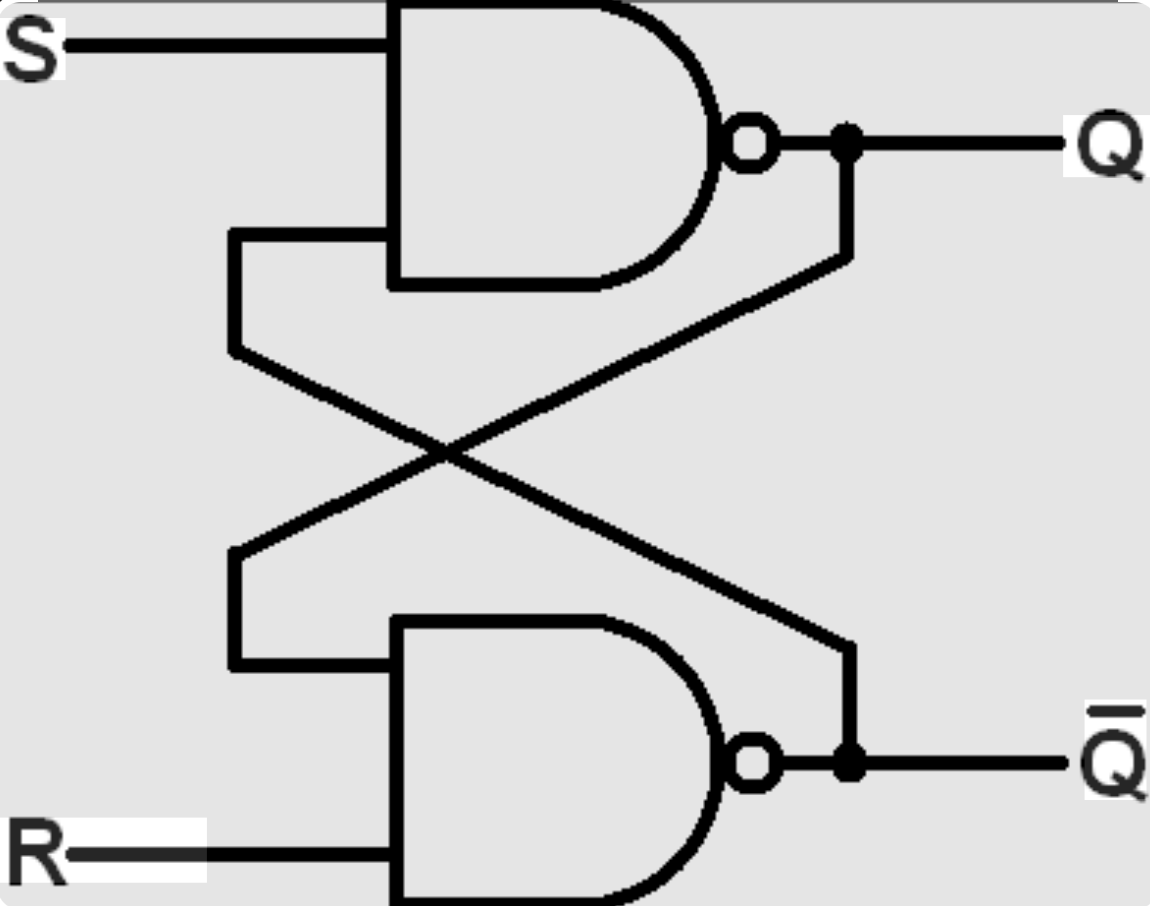} \end{center}
	Let \(\SR(w,1) = \G_1(w_1)\) and \(\SR(w,2)= \G_2(w_2)\)
 where\begin{itemize}
\item \(\G_1\) and \(\G_2\) are nand-gates with gate delay \(n\)
\item and  each \(w_i = w_i(w)\) where:
	\begin{itemize}
		\item	\(w_1(\ess)=\ess\) and \(w_1(w.a) = w_1(w).((a)_1,\G_2(w_2))\),
\item and \(w_2(\ess)=\ess\) and \(w_2(w.a) = w_2(w).((a_2,\G_1(w_1))\). 
\end{itemize}
\end{itemize}

The composition used to construct \(\SR\) is basically a function form
of the 
\emph{concurrent product} of state machines
\cite{yodaikenlarge,hartmanis}.

Claim: \(\SR\) is a SR latch with latch time \(3n+3\). \\
Proof:\\
Note that
	\(\held(w,1,b) = \held(w_1,1,b)\)
	and \(\held(w,2,b) = \held(w_2,1,b)\). These are just passed through 
and are obviously true for \(w=\ess\) and if true for \(w\) must
hold for \(w.a\) by the inductive hypothesis and the definitions
of \(w_i\).

It follows that:
\(\held(w,1,0) \geq n \) implies \(\held(w_1,1,0)\geq n\) which implies
that \(\G_1(w_11 = 1\). Now note for any \(k\), 
\(\held(w,1,0) \geq n+k\) implies \(\held(w_2(w),1,1)\geq k\). Proof by
induction on \(k\). For \(k=0\) there is nothing to prove. Suppose the
implication holds for \(held(w,1,0)=n+k\)  and consider
what happens if \(held(w.a,1,0)=n+k+1\).
 Since we know \(\G_1(w_1(w))=1\) if \(\held(w_2(w),1,1)=n+k\)
 it follows that
\(\held(w_2(w.a),1,1) = k+1\). So setting \(k=n\) \(\held(w,1,0) \geq 2n\) tells
us that \(\held(w_2,1,1)\geq n\) and given \(\held(w,2,1) \geq 2n\) we
know \(\G_2(w_2)=0\). Similar arguments take us to 
\(held(w,2,1)\geq 3n\) implies \(held(w_1(w),2,0)\geq n\). So 
for \(a=(1,1)\) \(\G_1(w_1(w.a))=1\). The case for inputs \((1,0)\) is
parallel. 

\section{More granular approaches \label{sec:alternatives}}
The model presented above is particularly simple. One addition might
be to specify hysterisis for the gates. This would allow a tighter
bound on timing for the latch. 
\[SinceStable(\ess,k) = 0\]
\[SinceStable(w.a,k) = \begin{cases}
	k&\text{if }\held(w,1,0) > n\text{ or }\held(w,2,0) > n\\
	&\text{or }\held(w,1,1) > n\text{ and }\held(w,2,1) > n\\
	k-1&\text{if }SinceStable(w.k)>0 \\
	&\text{and }\neg(\held(w,1,0) > n\text{ or }\held(w,2,0) > n)\\
	&\text{and }\neg(\held(w,1,1) > n\text{ and }\held(w,2,1) > n)\\
0&\text{otherwise }\end{cases}\]

Then \(SinceStable(w,k) >0\) implies \(\G(w.a)=\G(w)\).

Alternatively, in some circumstances these circuits could be represented
with inputs and outputs over a bigger range: say \(\{0,0.01,\dots 4\}\)
Or perhaps adding an unknown value would be useful.

\printbibliography
\end{document}